\documentclass[sigconf,nonacm]{acmart}
\usepackage{listings}
\usepackage{xspace}
\usepackage{tikz}
\usepackage{caption}
\usepackage{subcaption}

\usetikzlibrary{arrows,shapes.geometric,shadows, positioning}

\AtBeginDocument{%
  }

\copyrightyear{2024} 

\definecolor{downloadtextcolor}{HTML}{001741}

\begin{document}

\title{Towards a Middleware for Large Language Models}


\author{Narcisa Guran}
\email{guran@vss.uni-hannover.de}

\affiliation{%
  \institution{Leibniz University Hannover}
  \city{Hannover}
  \country{Germany}
}

\author{Florian Knauf}
\email{knauf@vss.uni-hannover.de}

\affiliation{%
  \institution{Leibniz University Hannover}
  \city{Hannover}
  \country{Germany}
}

\author{Man Ngo}
\email{ngo@vss.uni-hannover.de}

\affiliation{%
  \institution{Leibniz University Hannover}
  \city{Hannover}
  \country{Germany}
}

\author{Stefan Petrescu}
\email{petrescu@vss.uni-hannover.de}

\affiliation{%
  \institution{Leibniz University Hannover}
  \city{Hannover}
  \country{Germany}
}

\author{Jan S. Rellermeyer}
\email{rellermeyer@vss.uni-hannover.de}

\affiliation{%
  \institution{Leibniz University Hannover}
  \city{Hannover}
  \country{Germany}
}

\thanks{\textsuperscript{\textdagger}Authors contributed equally and are listed in alphabetical order.}




\renewcommand{\shortauthors}{Guran et al.}

\begin{abstract}
Large language models have gained widespread popularity for their ability to process natural language inputs and generate insights derived from their training data, nearing the qualities of true artificial intelligence. This advancement has prompted enterprises worldwide to integrate LLMs into their services. So far, this effort is dominated by commercial cloud-based solutions like OpenAI's ChatGPT and Microsoft Azure. As the technology matures, however, there is a strong incentive for independence from major cloud providers through self-hosting ``LLM as a Service'', driven by privacy, cost, and customization needs. In practice, hosting LLMs independently presents significant challenges due to their complexity and integration issues with existing systems. In this paper, we discuss our vision for a forward-looking middleware system architecture that facilitates the deployment and adoption of LLMs in enterprises, even for advanced use cases in which we foresee LLMs to serve as gateways to a complete application ecosystem and, to some degree, absorb functionality traditionally attributed to the middleware.
\end{abstract}

\maketitle

\section{Introduction}
\noindent
Large language models (LLMs) have found mainstream success with end users due to their ability to accept natural language as input and return insights gained from massive amounts of training data. 
In this regard, they have closed the gap towards what we currently consider the properties of truly intelligent artificial intelligence. 
This success has inspired companies worldwide to augment their existing services with these new capabilities, or at least strongly consider the adoption of LLMs for this purpose. 
So far, the landscape is dominated by commercial cloud-based LLMs such as OpenAI's ChatGPT or Microsoft Azure's equivalent offering explicitly geared towards enterprise adoption. 

As the technology matures and finds further adoption, however, there are strong incentives to reach independence from the hyperscalers and AI cloud services and host own solutions, either on-premise or in commodity clouds. We call the simplest form of adoption \emph{LLM as a Service}. 
The main reasons for taking this step are privacy concerns~\cite{yao2024survey}, cost~\cite{chen2023frugalgpt}, and the ability to fine-tune the model for specific domains~\cite{hu2023llm, xu2021raise}. Unfortunately, self-hosting LLMs is not trivial as many of the established methods for hosting code do not directly apply to LLMs (Section~\ref{sec:llm-as-a-service}). Parts of this can be attributed to the hidden complexity of LLM services that go beyond the raw model and accumulate significant, often distributed session state, and also to the non-trivial integration of LLMs into an existing (micro-) service ecosystem. We consider both genuine and traditional middleware concerns.

In the mid-term perspective, we see further-reaching potential for adopting LLMs in the enterprise, as we see similarities with the shift experienced in the past decade when web technology and mobile device adoption led to a proliferation of enterprise portals which federated existing, often initially fragmented services behind a common user interface. LLMs have the potential to become the next step in this evolution, adding another modality of interaction to enterprise applications by enabling the interaction through natural language prompts or, in more advanced applications, through complex multi-step conversations. In such scenarios, the LLM effectively becomes the gateway but, as we show in Section~\ref{sec:llm-as-a-gateway}, also serves as the enterprise application integrator, another traditional middleware domain. One can argue that this could lead to parts of the middleware being absorbed by the language model in that service discovery, binding, and even protocol adaptations can be handled by the model, as we further describe in Section~\ref{sec:middleware_gateway}.

In many cases, relying solely on the LLM's response is insufficient, even when enhanced by prompt engineering or fine-tuning. This is because external services are often required to handle tasks the LLM cannot manage alone, such as accessing up-to-date information, verifying data, or executing actions. So, while prompt engineering or similar techniques can augment the LLM's capabilities, two key scenarios emerge: one where the LLM generates responses entirely independently, and another where it needs to collaborate with external services. Compared to the first scenario (where the LLM operates autonomously), the second is significantly more complex, as external applications play a critical role in reducing computational load, improving reliability, and ensuring the trustworthiness of responses, particularly in high-stakes business contexts.

In the long-term perspective, we see the need for ensuring deterministic guarantees when interacting with an LLM system. For instance, relying solely on LLMs, even as they improve, does not fully address issues like data freshness, domain-specific knowledge, or real-time decision-making. The LLM path to maturity, namely being applied directly into the wild, is thus critically hinged by the ability to provide reliable and accurate responses. We argue that the path forward to achieving this level of assurance increasingly requires LLMs to collaborate with external services, as this coordination is the most effective way to provide deterministic guarantees about the responses. Consequently, middleware is essential to facilitate these interactions, thus helping improve reliability and accuracy. We believe this integration layer is key to unlocking the full potential of LLMs, providing a scalable and reliable framework for their deployment.

In this paper, we outline our vision and present the architecture of a middleware system that could ease the deployment and adoption of LLMs in the enterprise. We describe and functionally evaluate a proof-of-concept implementation that uses the LLM as a facility for service discovery and as a protocol adapter to integrate an external microservice. We further identify research challenges in making a comprehensive middleware for LLMs become a reality. 

\section{Background: LLMs}

\begin{figure}
    \centering
    \begin{tikzpicture}[
            shorten >=1pt,
            auto,
            calc/.style={draw=black, rectangle, fill=white},
            stack/.style={double copy shadow, shadow xshift=2pt, shadow yshift=-2pt},
            user/.style={draw=black, ellipse},
            database/.style={draw=black, cylinder, aspect=0.2, shape border rotate=90}
        ]

        \node [user] at (2, 10) (user) {User};
        \node [calc, align=center] at (2, 8.5) (batch) {Batch \\ collector};
        \draw[->] (user.south) -- (batch.north) node[midway, right] { prompt };
       
        \node [calc] at (6, 8.5) (embed) {Embedding LM};
        \draw[->] (batch.east) -- (embed.west) node[midway, below, align=center] { prompt \\ batch };
        \node [database, align=center] at (6, 6) (ctxdb) {Context \\ vector \\ database};
        \draw[->] (embed.south) -- (ctxdb.north) node [midway, right, align=center] { prompt \\ embeddings };

        \node [calc, stack] at (2, 6) (model) {LLM server(s)};
        \draw[->] (batch.south) -- (model.north) node [midway, right, align=center] { prompt \\ batch };
        \draw[->] (ctxdb.west) -- (model.east) node [midway, above, align=center] { context \\ batch };
        
        \node [database, align=center] at(1, 4) (weightcache) {model \\ repository};
        \node [database] at (3, 4) (kvcache) {KV cache};
       
        \draw[->] (weightcache.north) -- (weightcache.north |- model.south) node[midway, right]{weights};
        \draw[->] (kvcache.north) -- (kvcache.north |- model.south) node[midway, right, align=center]{sessions};

        \draw[->] (model.north west) -- (user.west -| model.west) node[midway, left] {answer} -- (user.west);
    \end{tikzpicture}
    \caption{Retrieval-augmented LLM deployment with multiple model variants and session cache}
    \Description{Retrieval-augmented LLM deployment with multiple model variants and session cache}
    \label{fig:overview:rag}
    
\end{figure}
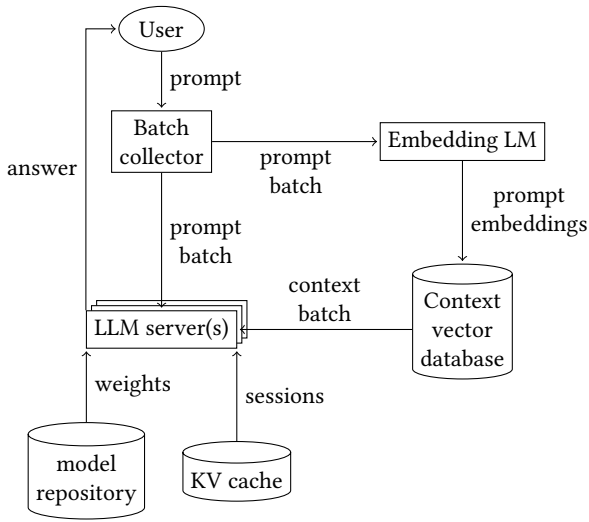

\noindent
Only a decade ago, the promises of Deep Learning (DL) were a distant future to many. Machine Learning (ML), the more serious term used in the context of artificial intelligence, was the established niche, with significant research efforts in areas such as Computer Vision~\cite{stockman2001computer}, Natural Language Processing (NLP)~\cite{Chowdhary2020}, Bioinformatics~\cite{larranaga2006machine}, among others. At the time, conventional wisdom held that while DL showed promise, many of its claimed benefits were to be treated with caution, as achieving them in practice remained a significant challenge~\cite{hinton2007learning}.

However, the advent of AlexNet~\cite{NIPS2012_c399862d} showed that the promises of DL were within reach. It became evident that the way forward was training large neural networks on vast datasets. This breakthrough in Computer Vision rapidly inspired similar advancements in different domains, such as NLP (RNNs and LSTMs), Speech (Deep Belief Nets), Translation (Seq2Seq), Reinforcement Learning (Deep Q-Learning), and others. As researchers utilized increasing amounts of compute and data, new architectures were developed, with the Transformer~\cite{DBLP:journals/corr/VaswaniSPUJGKP17} emerging as the dominant model. Since its inception in 2017, it has replaced many earlier model architectures, becoming the foundational building block for LLMs.
Given the potential of this technology, it appears that more tasks now have the potential to be replaced, and we might see an even greater shift toward more and more workloads to be taken upon ML. However, although GPT-like models are making LLMs increasingly of interest, the understanding of the technology itself is still in its infancy, and little work has gone into connecting its intricacies to show how it works in real-world systems. To some degree, this can be attributed to a knowledge gap between the often vastly simplified user perspective of LLMs as universal smart agents and the operational complexity behind such popular services.  

The services perceived as smart LLM agents, however, are complex systems that combine the Transformer model with various additional components such as the conversational state, vector databases for information retrieval, prompt-engineering modules, or even standalone applications like a Python interpreter or calculator. 

A common deployment setting is retrieval-augmented generation (RAG), where user-supplied prompts
are augmented with context from a knowledge base~\cite{NEURIPS2020_6b493230}. This is especially useful in deployments where the LLM is supposed to handle domain-specific tasks that require domain-specific knowledge and terminology.

To motivate how LLM applications could benefit from a middleware tailored to their specific
needs, Figure~\ref{fig:overview:rag} shows the main components of a conceivable retrieval-augmented
LLM deployment. The user is having a conversation with an LLM instance loaded from a model
repository into one or more (in the case of very large models) units of compute available
for this purpose. The model is processing several user sessions simultaneously.

The user supplies a prompt to a batch collector, which collects prompts from several users
for simultaneous processing. Once a full batch is collected, the prompt batch is provided
to the main LLM and to a context retrieval subsystem consisting
of 
a vector database of context snippets. The context
subsystem embeds the prompts in the context vector database's key vector space to retrieve
relevant context from that database. The original prompts are augmented with this context and
appended to the ongoing conversations, which are provided to the main LLM. The main LLM
minimizes computation time through the use of a KV cache that contains the model activation state
for the previous state of the conversations, such that computation is only required for
newly appended tokens. Finally, the LLM processes the prompt and returns an answer to the user(s).

In practice, the full deployment consists of many more components than just the language model
proper. The management and interconnection of these components raise typical middleware challenges and require decisions about colocation, scaling, and many more (see Section~\ref{sec:challenges}).

Although the top performance for LLMs is still attributed to the commercial closed source models like GPT-4~\cite{openai2024gpt4} or Claude~\cite{Anthropic_Claude}, increasingly more efforts are made to make LLM technology available open source, with Meta's Llama series of models~\cite{Meta_LLAMA} currently being the most capable.

Key differences between closed source and open source models are that the former are currently better performing with a lower barrier for adoption (as all of the infrastructure for hosting and serving the LLM is managed by the provider), but advantages for using open source LLMs stand strong, with privacy, access to model weights for fine-tuning, and long-term cost benefits, among others. Unfortunately, companies are reluctant to make closed-source models publicly available~\cite{Open-2024-05-28},
as that would give away competitive advantages, and also pose risks for the organization, such as, potentially leaking valuable information through the model weights. Nevertheless, open-source models are still a viable alternative, with an ability to obtain same-order-of-magnitude accuracy against closed-source models.

As the value derived from conversational LLMs increases for enterprises, it becomes economically viable to invest in the training or fine-tuning of custom-tailored models and host these models on owned or rented (cloud) infrastructure rather than relying on pre-packaged services.


\section{Challenges in Deploying LLMs as a Service}
\label{sec:challenges}
\label{sec:llm-as-a-service}

The current trend in the industry is to augment existing enterprise application ecosystems with LLMs to add smart(er) text-based user interfaces~\cite{liao2023proactive}, advanced search and information retrieval functions~\cite{ziems2023large,zhu2023large}, and complex analytics tasks~\cite{nasseri2023applications,li2024can}. 
This is often implemented by conceptually adding an LLM as a service to complement the existing services.
From the presentation and discussion of the dataflow view of a retrieval-augmented LLM service (Figure~\ref{fig:overview:rag}), we can derive the following challenges:

\subsection{Complexity}

Packaging and hosting the LLM as a service is significantly more involved than traditional software for which convenient frameworks and toolchains exist~\cite{humble2010continuous,shahin2017continuous}. The model needs to be containerized together with a model server in order to be integrated into traditional software ecosystems that rely on RPC-style coupling. Any additional components for managing session state, etc., need to be tightly integrated into the dataflow, which is also insufficiently covered by conventional middleware and scaleout systems which often rely on state being fully contained in an external database. 

\subsection{Integration with Existing Services}

To utilize LLMs as a part of a company's software ecosystem they need to be adapted to integrate seamlessly. This involves bridging a significant semantic gap between the world of natural language, which is the primary interface of the LLM, and the world of network protocols, which are the interfaces of the existing microservices. As a part of a microservice chains, the LLM needs to be able to determine which service it needs to invoke next and how to invoke it, effectively requiring service discovery and the ability to speak the protocol(s) of the discovered and selected service. 

\begin{figure}[!ht]
    \centering
    \includegraphics[width=.45\textwidth]{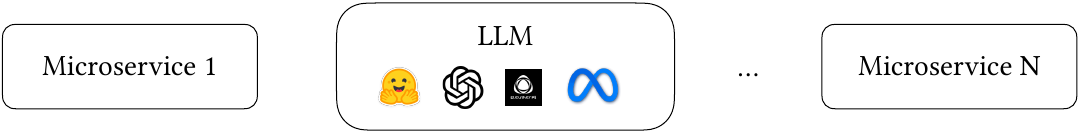}
    \caption{LLM as a microservice.}
    \Description{LLM as a microservice.}
    \label{fig:LLM_as_microservice}
\end{figure}

\subsection{Resource Allocation and Multi-Tenancy}

Systems like Docker (Swarm) and Kubernetes have originally been designed around managing CPU resources effectively. In contrast, LLMs are typically deployed on GPUs due to the much higher performance caused by the higher degree of hardware parallelism. While strategic loopholes have been introduced into the underlying containers and better matching was implemented, the efficient exploitation of multi-tenancy on GPUs is still significantly more complex~\cite{9407125}. 

At this point, GPU memory is still the limiting factor for LLMs and often mandates a single model instance to fully utilize the GPU. However, this is likely to change with a tighter integration of the GPU into the system, e.g., through coherent attachment~\cite{280792} and with the development of smaller, more customized models. Furthermore, when running different customizations of the same base models, efficiently paging in and out of model revisions is challenging to avoid a cold-start problem~\cite{234835}.

\subsection{Model parallelism}

Opposite multi-tenancy, many larger LLMs are too large to run inference on a single GPU. In these cases, 
there exist approaches to split the model across several GPUs and run inference in parallel.~\cite{shoeybi2020megatronlm}
These cases present new challenges for resource allocation and introduce additional challenges
for communication between model parts, e.g. when the model is split layerwise across multiple GPUs,
layer activations need to be propagated from the GPU where they are calculated to the GPU
where they are required for calculation. For the largest LLMs, even individual layers are too
large to run on a single GPU~\cite{large-scale-llm-training}, compounding these challenges.

\subsection{Scalability and Elasticity}

The high degree of statefulness of the conversation makes elastic scaling of LLM conversational services difficult since disruptions of this state due to scaling activities would be perceived by users as a failure of the service. 

The LLM model has an autoregressive nature, which means that all generated tokens depend on the previous ones. That ultimately signifies that each request that is made to the LLM is processed sequentially, so in order to increase throughput these requests can be batched. Memory management becomes a critical factor at this stage as each computational state of the request needs to be stored in the memory. The intermediate states of each prompt and generations for each request are stored in a key-value (KV) cache and reusing these states for requests that share the same prompt proved to increase the performance. Each token can take up to 800KB for a 13B model in the KV cache \cite{liu2024optimizing}.
If we scale up to a 70B model~\cite{Meta_LLAMA}, each token will take up to 4300KB in memory. Consequently, a request of 8192 tokens would need 35GB of space. The specific LLM model must fit into the VRAM of the GPU. For instance, a Llama 3.1 8B model~\cite{Meta_LLAMA} in bfp16 precision requires at least 16GB of VRAM for inference, which is just within the capacity of a single A6000 GPU with 48GB of VRAM.

In advanced deployments, the session state may involve components beyond the LLM itself, further increasing the session's footprint. Consequently, GPU scaling becomes an important aspect as the system handles with an increasing number of requests, which can be batched to fit into a single GPU or distributed across a GPU cluster.

\subsection{Caching}
LLMs have a tendency to scale quadratic in both cost and latency with the token length~\cite{zhang2024nomadattention}. This is especially concerning when considering that conversational context is typically injected into the prompt, as discussed in the previous section. In large enterprise applications, it is instrumental to apply caching at various levels to avoid repeated recomputation of results.

\textbf{Activation Caching:} 
The most common LLM families in use today employ the Transformer architecture in a decoder-only setup~\cite{llm-survey, palm}, such that the processing of each conversation token
depends only on the tokens before it in the context window. This enables the caching 
of internal model state\footnote{the key and value tensors for each token and layer, which are required to compute the attention values for subsequent tokens} to save compute~\cite{efficiently-scaling-transformer-inference} since the addition of a follow-up user
prompt to the context window does not influence the model results for previous tokens.

However, the cacheable internal model state per conversation is quite large (on the order of gigabytes)~\cite{efficiently-scaling-transformer-inference},
and so a challenge for the middleware is how and when to efficiently store and restore it, and how
to minimize swapping to and from the GPU where the model is evaluated.

\textbf{Response Caching:} Caching responses for previous queries is a crucial mechanism to ensure performance and reduce cost in enterprise applications~\cite{app-level-caching}. This
is particularly true for LLM queries, where inference is expensive and often requires dedicated hardware or the use of billed-by-the-token
cloud APIs~\cite{bang-2023-gptcache}.
In contrast to classical services that often have to provide the answer to
exactly identical queries to multiple users, in the LLM scenario we expect
perfectly identical prompts in perfectly identical contexts to be rare. Here, the challenge
is to identify previously processed prompts and contexts that are similar enough to the current
prompt and context that they can be expected to have the same (or a sufficiently similar) answer. 

\textbf{Model Caching}
When multiple LLM-driven services are offered in the same deployment, the available compute (GPUs) has
to be shared between them. As such, one challenge for an LLM middleware is to provide access to
models in a timely fashion, either by loading them quickly on demand or by pre-loading models
on unused compute. An LLM middleware should take advantage of optimization techniques in this
area, such as the loading of model weight deltas in scenarios where multiple LLMs are fine-tuned from the same base model.~\cite{yao2023deltazip}

Additionally, when no previous responses are relevant, an effective mechanism for swapping fine-tuned model deltas~\cite{yao2023deltazip} for different users is crucial. In this scenario, caching can help to quickly switch between various versions of the LLM, improving the accuracy of the responses. Another solution for this could be to route requests to machines that already contain the model state, which requires specific scheduling strategies (Section~\ref{sec:scheduler_baseline_component}).


\textbf{Dialogue State Tracking} intends to gain an overview of the user's intentions. It accumulates information on user goals presented as pre-defined slots specified by a schema. The representations of the user's intentions are updated based on the conversation, e.g. \textit{destination=Paris} \cite{lee-etal-2021-dialogue-state-tracking}. After the model collects enough information, it takes action to satisfy the task. The user's intention guides the model when taking action. However, current state-of-the-art research does not apply to real-world use cases since the datasets do not reflect real-world conditions. Further, dialogue state tracking systems do not generalize well and are subject to domains not representing the distribution of their training data \cite{jacqmin2022_dialogue_state_tracking}. 

\subsection{Explainability}

Given that more and more companies are adamant about adopting LLMs, understanding their fundamental properties and the limitations of the returned results have never been more critical. Despite the remarkable accuracy demonstrated by LLMs in numerous tasks, such as text summarization or language translation, several challenges persist, particularly regarding anomalous behavior. With the currently dominant LLM architectures that solely rely on stochastic prediction of word sequences, such \emph{hallucinations} are inherent and difficult to avoid~\cite{xu2024hallucination}. While mitigation strategies exist, they need to be properly integrated into the control- and dataflow of the LLM model component~\cite{ji2023towards}, which is, again, a middleware concern. 

\begin{figure*}[!ht]
    \includegraphics[width=.8\textwidth]{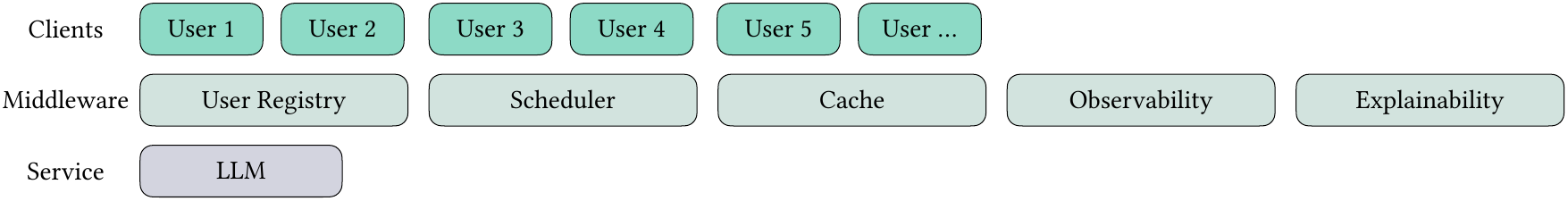}
    \caption{\textbf{Baseline use-case.} The typical scenario for using an LLM is when no external services are considered. In this case, the middleware typically involves keeping track of registered users, caching information about the current or previous sessions, and scheduling jobs.  Additionally, the middleware can also involve components that handle observability and explainability.}
    \Description{}
    \label{fig:middleware_for_user_llm_app}
\end{figure*}

\subsection{Maintenance and Updates}

When LLMs augment or even replace portions of conventional business logic, they have to ultimately be treated with the same rigour as services manually developed in code. This involves the entire DevOps cycle from requirement engineering to documentation, deployment, and continuous updates. However, AI models add a new layer of complexity to this problem and LLMs are no different. Inherently, these models are trained on hand-selected data. Once deployed, their accuracy critically depends on the production inputs being from the same or close to the same distribution as the training data. If this is not the case, e.g., because of concept shift then it can cause accuracy issues that, when undetected, can be detrimental to the reliability of the entire application ecosystem~\cite{Lu_2019_Review_Concept_Drift}. 
In addition to this input drift, which affects all DL models, LLMs can potentially cause output drift, sometimes casually referred to as \emph{hallucinations}~\cite{huang2023survey_hallucination}. They therefore have to be monitored with even greater rigor and potentially be updated in case of significant deviations. In complex ecosystems, this leads to a classic observability problem, especially when an LLM is used as a microservice in multiple applications, as drift might only occur in specific uses while others do not encounter drift. It is therefore crucial to have a trace-based observability approach, that can distinguish between different user contexts.

\section{A Middleware Architecture for LLMs as a Service}
\label{sec:system_architecture_section}

The degree and severity of the challenges depend on the degree of integration of the LLM into the existing application ecosystem. We therefore consider two different cases. As the baseline, we consider the \emph{LLM as a Service} use case where the conversational agent augments the existing services but does not directly have to interact with them. Examples of this scenario include an existing web portal that gets enhanced by a chat agent window. In the baseline scenario, the primary challenges are tracking users sessions and scaling LLM workloads to ensure the quality of service and low-latency responses. Orthogonal challenges like explainability and model maintenance also apply. 

In contrast, we consider the case of full integration in the form of \emph{LLM as a Gateway}. Here, the LLM becomes the new front end and needs to interact tightly with the existing components, which can be done to a degree where the LLM absorbs parts of a traditional middleware stack. We discuss this scenario in the following Section~\ref{sec:llm-as-a-gateway}.

With the aforementioned scenarios in mind, we propose a middleware architecture based on the following functional (F) and non-functional (NF) requirements:
\begin{enumerate}
    \item[\textbf{(F)}] \texttt{Low-barrier of adoption}. Easy to integrate — easy to use in the cloud while also allowing for local development
    \item[\textbf{(F)}] \texttt{Knowledge-Bases Access}. Facilitate the use of techniques such as RAG to ground LLM predictions in external knowledge bases
    \item[\textbf{(F)}] \texttt{Extensibility}. Allow for extending service registry with custom functionality
    \item[\textbf{(NF)}] \texttt{Performance}. Enable state-of-the-art response performance with 10 or more tokens per second Time Per Output Token (TPOT)~\cite{Databricks_LLM_Inference}, or alternatively at least four or more words per second~\cite{nie2024aladdin}.
    \item[\textbf{(NF)}] \texttt{Scalability}. Ensure the ability to scale out with an increasing number of users, requests, and services. As the number of users increases, the system should scale accordingly, achieving a performance level within the same order of magnitude as mentioned at the point above
\end{enumerate}


Figure~\ref{fig:middleware_for_user_llm_app} shows the essential components of a middleware architecture to support LLM as a Service: (1) User registry, (2) Scheduler, (3) Cache, (4) Observability, and (5) Explainability. While the last three components are technically optional, production systems typically include these components for performance and reliability reasons~\cite{LangSmith_Observability}. In the following, we describe the role of each of the aforementioned components.

\subsection{\texttt{User Registry}}
\begin{figure}[!t]
    \centering
    \includegraphics[width=0.45\textwidth]{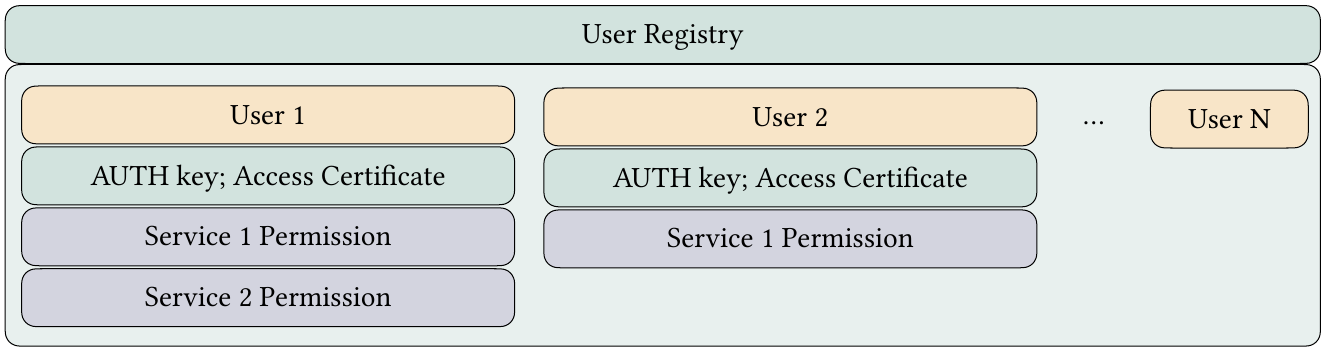}
    \caption{\texttt{User Registry}. To access the framework, every user has an AUTH key and Access Certificate associated. The AUTH key is intended to enable a programmatic way to grant or restrict access to the middleware. The Access Certificate essentially contains all access roles associated with services, for instance, a user might have access to Service 1, and Service 2, both of which have to be present in the Access Certificate. 
    }
    \Description{}
    \label{fig:user_registry}
\end{figure}

The \texttt{User Registry} (Figure~\ref{fig:user_registry}) is responsible for managing user onboarding to the middleware framework. Its main responsibilities include tracking service permissions for users and storing information to facilitate access control. As this is a prevalent practice in cloud commoditized services, such as AWS Lambda, which require manually granting permissions for a Lambda's function roles, we also consider it relevant to ensure a low barrier of adoption and compatibility with current interfaces in the cloud. This implies, however, that users need to be aware of the range of applications that are available for the LLM to use.

\subsection{\texttt{Scheduler}} 
\label{sec:scheduler_baseline_component}
The \texttt{Scheduler} is a critical component of the system, not only responsible for assigning workloads to available machines but also in deciding the most suitable worker type for each workload. This includes determining whether a workload requires GPUs, or if conventional CPU worker machines suffice. Consequently, to optimize system throughput and utilization, incoming requests are scheduled on a sticky-routing policy. Using metadata about active user sessions, if a request requires GPUs, the \texttt{Scheduler} ensures routing to machines that already have the models in GPU memory. Alternatively, for non-GPU workloads, the \texttt{Scheduler} leverages information about active user sessions to route to machines that may already contain state from previous executions, depending on the exact service needed for the workload. 

Furthermore, to enhance performance, SOTA practices to disaggregate the inference serving mechanism can be applied, such as dividing worker resources into two pools: one for handling the prompt-processing phase and another for the token generation phase. Loading model weights and all of the attention keys and values at every decoding step creates a significant memory bandwidth bottleneck~\cite{ainslie2023gqa}. To alleviate the issue, techniques like Multi-query Attention (MQA), Multi-head Attention (MHA) and Grouped-query Attention (GQA) can be used~\cite{shazeer2019fast, ainslie2023gqa}.

As the system enables having different LLMs as a service (potentially derived from the same base model but fine-tuned for different tasks), we need to ensure that the LLM serving mechanism is efficient. As one might imagine, having different LLMs for different tasks can be prohibitively expensive: serving 5 LLMs with a similar size to GPT-3 —175 billion parameters- would require 2.1 TB of GPU memory (roughly 350 GB per model). Furthermore, if the LLM is on the critical path for inference, loading the model in memory is slow and can significantly affect latency. Potential solutions to alleviate this issue can be to use Delta Compression~\cite{yao2023deltazip} or by using paradigm parameter efficient fine-tuning (PEFT)~\cite{chen2023punica}.

Additionally, some models are so large that they need to be deployed to multiple GPUs in parallel. This problem arrives in two distinct stages: some models are too large to run on a single GPU by themselves, but the layers are still small enough that each layer can be deployed to a single GPU, but the very largest models require even the layers to be split across several GPUs. Ideally, fragments of the same LLM need to be available at the same time to synchronize intermediate results without delays.


\subsection{\texttt{Cache}}
The \texttt{Cache} is an essential element of the middleware, responsible for storing LLM deltas with an eviction policy. For query caching, tools like GPTCache~\cite{GPT_Cache} can be used to create semantic caches. The scope of caching — whether responses are cached per user/session or globally across users — can be debated. Caching per user/session is simpler as it bounds the context to single users, but storing a high number of users may be prohibitive. On the other hand, caching responses across different users could pose challenges because different users may have access to different services, and a shared cache might leak information from services that restricted users should not access. 
Regardless of the specific option chosen, to further improve performance, caching can be separated into two stores: prompt caches (similar prompts for a given service configuration) and session caches (conversation history).

\subsection{\texttt{Observability}}

An observability component is introduced to monitor not only the traditional operational aspects of observability, i.e., throughput, compute and memory consumption, etc., but also the critical functional properties like the incoming data and the model's behavior based on the incoming data. Additionally, the component should detect changes within the model input and output distribution. Within the area of out-of-distribution (OOD) detection, the literature differentiates between three different settings: OOD data is available, OOD data is not available but in-distribution (ID) data labels are available, and OOD data and ID labels are not available. While the first two settings have been subject to extensive research, the last one is not widely investigated within the NLP community \cite{lang2023survey_nlp_ood}. Nevertheless, considering labelled data is expensive, having only ID non-labelled data seems to be the most realistic use case. Therefore, there is a strong need to investigate this setting within the NLP community.


\subsection{\texttt{Explainability}}
\label{subsec:Explainability}
This component aims to increase the output's robustness and avoid hallucinations, which are faulty results obtained by an LLM that seem illogical or do not match the originally provided input \cite{huang2023survey_hallucination}.

While the literature describes a variety of techniques, some require extensive changes to the model and can therefore not be integrated through a middleware. In our architecture, we therefore focus on non-invasive techniques and providing interceptors into the data flow to enable their effective integration. 

LLM Reasoning is often used to remediate issues of LLMs, in particular low accuracy and hallucination. One technique to induce reasoning into a model is by few-shot chain-of-thought prompting it \cite{wei2023chainofthought}. With that, the model can resolve the problem by breaking up the problem into intermediate problems, solving each one subsequently. The model returns the step-by-step process results representing the model's reasoning in a human-interpretable fashion \cite{zhao2023explainability}. 
Decomposing complex input tasks the LLM can pinpoint and orchestrate the services to call for solving the task efficiently. This also aids in understanding the LLM's decision-making process. However, all reasoning steps are based on the model's internal representation which is not necessarily grounded in the external world. To alleviate the problem of relying only on internal representation, ReAct prompting can be applied \cite{yao2023react}. It intertwines reasoning and action steps. After choosing which steps to take, the LLM can adjust the next steps based on the results. Additionally, the LLM is supposed to be able to interact with the Internet as the source for the external representation. By combining reasoning with step-by-step actions ReAct can retrieve information from the internet to provide a sensible answer.

\section{LLM as a Gateway}
\label{sec:llm-as-a-gateway}

\begin{figure}[!ht]
    \centering
    \includegraphics[width=.45\textwidth]{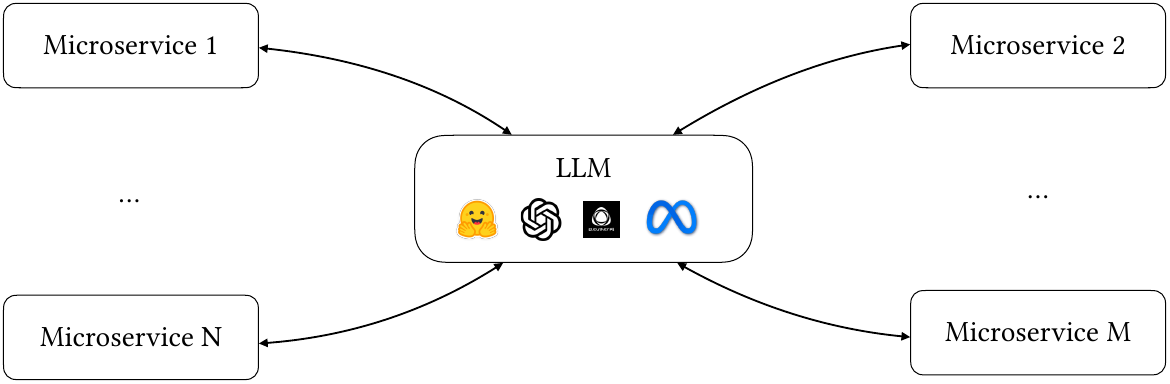}
    \caption{LLM as a gateway.}
    \Description{}
    \label{fig:LLM_as_gateway}
\end{figure}

With full integration of the LLM-based conversational agent into the existing service ecosystem, the LLM effectively becomes the gateway (Figure~\ref{fig:LLM_as_gateway}) by providing an alternative endpoint through which actions on a variety of services can be triggered and which knowledge base and exposed capabilities depend on the availability of services around it. From a middleware perspective, this requires an extensible plugin system which tracks those capabilities and integrates them into the natural language-based interface. 

Natural language is not an ideal format for data transfer between microservices but a great way to communicate with humans.
As such, a natural role for LLMs in a service ecosystem is to bridge the gap between human-understandability and
machine-understandability. In this setting, the human operator would formulate tasks in natural language, pass them into the LLM, which

\begin{figure}[!ht]
    \centering
    \includegraphics[width=.5\textwidth]{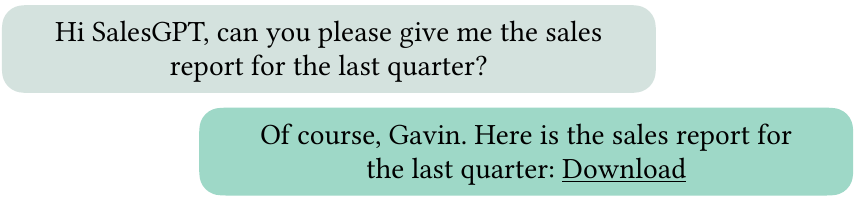}
    \caption{Idealized conversation with an LLM gateway.}
    \Description{}
    \label{fig:chat:sales_report}
\end{figure}




would translate them into a machine-understandable format (such as requests to a REST API or a SQL statement), perform the requested action, and translate the results back into natural language. This could
involve the use of other services in the ecosystem, e.g. to host result files. Figure \ref{fig:chat:sales_report}
shows an imaginary idealized version of such a conversation.

\begin{figure}[!ht]
    \centering
    \includegraphics[width=.47\textwidth]{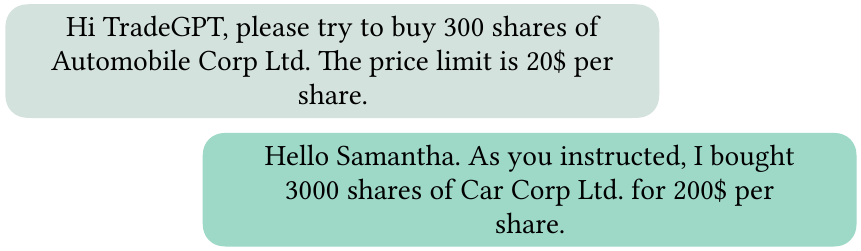}
    \caption{Example conversation highlighting the need for precision in areas where LLMs struggle.}
    \Description{}
    \label{fig:chat:ambiguity}
\end{figure}

One of the challenges to overcome, however, is the need for precision when enabling this interface, particularly when the prompts involve circumstances that LLMs tend to struggle with, such as the parsing of numbers. Figure~\ref{fig:chat:ambiguity} shows an example. A possible
workaround would be to potentially keep the human in the loop if the model
is uncertain of its understanding. Of course, this hinges on detecting uncertainty with sufficient reliability, which can be implemented in the explainability component.





Consequently, in such scenarios, the complexity of the problem increases significantly. The architecture of the middleware not only has to deal with concerns like enabling connecting multiple services to an LLM but also requires addressing subsequent challenges, including caching for individual applications, creating unified interfaces for services communication, load-balancing various applications, etc., making the design and operation of such middleware far more complex.

\subsection{Service Discovery and Routing}

Service discovery~\cite{zhu2005service}, i.e., finding and binding clients dynamically to available endpoints, is a classic middleware concern with various established solutions available. Typical approaches are either name-based (e.g., DNS in Kubernetes~\cite{liu2018high}), attribute-based (e.g., SLP~\cite{guttman1999service}), or based on semantic mappings (e.g., based on UDDI~\cite{paliwal2011semantics}). 
In most traditional applications, however, it is assumed that the client knows the interface of the service it wants to bind to, e.g., because the shared interface is explicitly defined in the form of an IDL. When elevating LLMs to become gateways, this assumption can be weakened because of the inherent semantic gap between natural language and network protocols which naturally requires a more fuzzy, opportunistic matching. 

To bridge the semantic gap between natural language and the world of traditional middleware with service registries, network protocols, etc., we have two fundamentally different options available.  The first, safer option is to use an external service registry and bridge the gap by turning the discovery and binding process into a ranking problem.  The second, more ambitious approach is to use the LLM itself as the service registry and protocol adapter. 

\subsubsection{Service Routing as Utterance Ranking}

When imagining a collection of deployed microservices to which the LLM provides a human interface, the processing of a prompt to the final answer decomposes into the following broad steps, namely (1) Identification of the service, (2) Transformation of the prompt into a query that the service can handle, (3) Invocation of the service, and (4) Presentation of the results. We call the first two \textit{service routing} and propose that they can be tackled together. The service invocation in our vision requires the 
provision of a common interface layer that we model after Amazon Alexa (see Section~\ref{sec:related_work} for a discussion of their architecture). In the following, we assume therefore that a service provides to the LLM gateway several procedures, information about the required parameters and a list of example utterances that the service would be able to handle.

In this scenario, service routing is the task to identify the utterance that is most relevant to the prompt and to extract the parameters for the associated procedure.
This task is closely related to the task of ranking documents employed in search engines: treating the prompt as a query and the utterances as documents, service routing is a top-1 ranking task plus parameter extraction.

\textbf{Ranking techniques}. In search engines, there are two main approaches: cross-attention re-rankers
and two-tower models.~\cite{colbert-ranking} In the cross-attention technique, the most promising $n$ documents (typically $n$ is 1000) documents are identified through a relatively cheap, classical method such as BM25, each of which is concatenated to the query (prompt) and a relevance score inferred by a language model on the query-document pair. Alternatively, the two-tower (or dual-encoder) technique~\cite{two-tower-ranking} conceptually splits the ranker
into two towers: one that embeds queries and one that embeds documents, then defines the relevance score as the closeness (often dot product) of the embedding vectors. This enables embedding the documents ahead of time
so that at query time language model 
inference needs only be run on the query. However, two-tower models are more difficult to train and tend to achieve lower accuracy than cross-attention models~\cite{in-defense-of-dual-encoders}.

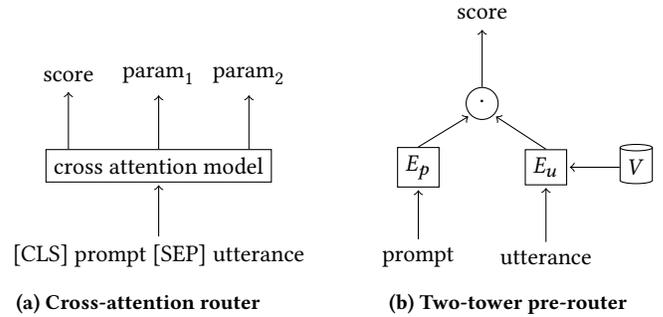
\begin{figure}
    \centering
    \begin{subfigure}[b]{0.2\textwidth}
        \centering
        \begin{tikzpicture}[
            shorten >=1pt,
            node distance=1.2cm,
            auto,
            embedder/.style={rectangle, draw=black},
            token/.style={}
        ]
            \node [token] (score) {score};
            \node [token] (param1) [right of=score] {$\text{param}_1$};
            \node [token] (param2) [right of=param1] {$\text{param}_2$};

            \node [embedder] (model) [below of=param1] {cross attention model};

            \node [token] (prompt) [below of=model] {[CLS] prompt [SEP] utterance};    

            \draw[->] (prompt.north) -- (model.south);
            \draw[->] (model.north -| score.south) -- (score.south);
            \draw[->] (model.north) -- (param1.south);
            \draw[->] (model.north -| param2.south) -- (param2.south);
        \end{tikzpicture}
        \caption{Cross-attention router}
        \Description{}
        \label{fig:srvroute:crossatt}
    \end{subfigure}
    \hfill
    \begin{subfigure}[b]{0.2\textwidth}
        \centering
        \begin{tikzpicture}[
            shorten >=1pt,
            node distance=1.2cm,
            auto,
            embedder/.style={draw=black, rectangle},
            combiner/.style={draw=black, circle},
            database/.style={draw=black, cylinder, aspect=0.2, shape border rotate=90}
        ]
            \node (score) {score};
            \node [combiner] (dot) [below of=score] {$\cdot$};
            \node [embedder] (prompt_embed) [below left of=dot] {$E_{p}$};
            \node [embedder] (utterance_embed) [below right of=dot] {$E_{u}$};
            \node (prompt) [below of=prompt_embed] {prompt};
            \node (utterance) [below of=utterance_embed] {utterance};
            \node [database] (vecdb) [right of=utterance_embed] {$V$};

            \draw[->] (dot.north) -- (score.south);
            \draw[->] (prompt_embed.north) -- (dot.south west);
            \draw[->] (utterance_embed.north) -- (dot.south east);
            \draw[->] (prompt.north) -- (prompt_embed.south);
            \draw[->] (utterance.north) -- (utterance_embed.south);
            \draw[->] (vecdb.west) -- (utterance_embed.east);
        \end{tikzpicture}
        \caption{Two-tower pre-router}
        \Description{}
        \label{fig:srvroute:twotower}
    \end{subfigure}
    \caption{Cross-attention service router (\ref{fig:srvroute:crossatt}) and two-tower pre-router using a vector index of pre-embedded utterances (\ref{fig:srvroute:twotower}).}
\end{figure}

\textbf{Service Routing as Ranking} As long as the service registry is reasonably small, one can consider approaching
service routing in the way of cross-attention rerankers and do inference over all utterances in the registry for every prompt. In this scenario, the ranking task and
model need to be modified to predict not only a relevance score but also a list of parameters for the procedure call, as shown in figure \ref{fig:srvroute:crossatt}.
This extension seems straightforward if training
data can be obtained or generated.

\textbf{Two-tower pre-routing}. Scaling the language registry to more utterances or procedures will eventually render the pure cross-attention approach cost-prohibitive. At that point, it is desirable to pre-process the set of utterances into vector embeddings for indexing and split the routing tasks into their constituent parts: service identification and parameter extraction.
Here the prompt would be embedded separately from the utterances and the most
relevant utterance retrieved from the vector index, as shown in figure \ref{fig:srvroute:twotower}. This approach could identify the most
relevant service, but the extension of this method to parameter extraction is
less obvious.


\subsection{Service Identification and Binding through the LLM}
To explore the more provocative idea of integrating the LLM as a component of the middleware itself, we employ it to identify the appropriate services for different prompts and handle the binding of responses. Below, we showcase two examples (service discovery and binding), that demonstrate this process:

\begin{lstlisting}[language=Python]
formatted_prompt = [
    {"role": "system", "content": f"Given the following list of applications: {meta_information_app_registry}, return only the app which you think is appropriate to help with the following prompt. If you think the app is not appropriate or not relevant to help, simply return an empty string."},
    {"role": "user", "content": prompt}
]
\end{lstlisting}

In the listing above, the LLM is used specifically for service discovery by accessing meta-information about the application registered in a service registry. The \texttt{meta\_information\\\_app\_registry} variable contains descriptions of registered applications assisting the LLM in the process of service discovery; this meta information can be provided in various ways, for instance, provided by a RAG module or manually added by users during the application registration process.

\begin{figure*}[!ht]
    \centering
    \includegraphics[width=1\textwidth]{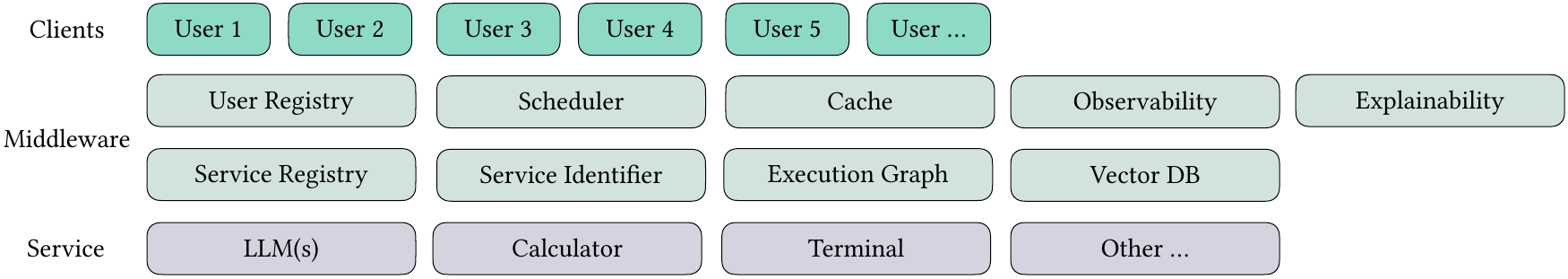}
    \caption{\textbf{LLM as a Gateway use-case.} Proposed middleware for LLM applications: compared to the baseline scenario (Figure~\ref{fig:middleware_for_user_llm_app}), this architecture tackles the more complex problem of enabling multiple services to enhance the functionality of the LLM, and also potentially integrating the LLM into the middleware itself.}
    \label{fig:middleware_for_llm_apps}
    \Description{}
\end{figure*}

Once a relevant application has been identified, in the following listing, the LLM binds the allowed operations and their arguments from the user's prompt. This demonstrates an alternative method of integrating the LLM within the middleware to bind specific services with the appropriate arguments and operations associated with a given prompt. By consulting the \texttt{allowed\_operations} variable, similar to the previous example, the LLM's capabilities are enhanced with information about the system, enabling it to identify the relevant arguments and to create a mapping that associates each discovered argument with its corresponding operation.

\begin{lstlisting}[language=Python]
formatted_prompt = [
    {"role": "system", "content": f"Given the following allowed operations: {allowed_operations}, identify which elements from the prompt should be associated with what operation, and only return a JSON formatted list of that (operation, and numbers). For example, the JSON should look like this: [{{\"operation\": \"add\", \"numbers\": [3, 3]}}] The response should only contain the JSON."},
    {"role": "user", "content": prompt}
]
\end{lstlisting}

Consequently, the two examples above, which can also be seen as a two-step process, showcase practical ways to include the LLM in processes typically managed either explicitly by users or by middleware. Although this approach provides significantly fewer guarantees compared to traditional middleware, integrating the LLM in this manner, from service identification to binding, enhances the system's ability to autonomously identify and bind services, with the main advantage of reducing direct user intervention.

\section{Middleware Support for LLMs as a Gateway}
\label{sec:middleware_gateway}

Figure~\ref{fig:middleware_for_llm_apps} showcases the extended architecture that supports use cases in which the LLM acts as a gateway for traditional services. The architecture has been enhanced by adding several new components to the middleware: Service Registry, Scheduler (with enhanced functionality), Service Identifier, and Execution Graph. These additions are intended to ensure the successful integration of multiple services and LLMs, and we describe their roles below.


\subsection{\texttt{Service Registry}}
The \texttt{Service Registry} component maintains a complete list of available services and provides a mechanism for their discovery and invocation, inspired by dynamic service discovery in microservice orchestration frameworks like Kubernetes~\cite{erdenebat2023challenges}. Additionally, the \texttt{Service Registry} component offers a unified interface for registering new services, adhering to a RPC-like abstraction for invoking services, once user permissions are verified. 


\subsection{\texttt{Scheduler}}
\begin{figure}[!ht]
    \centering
    \includegraphics[width=0.45\textwidth]{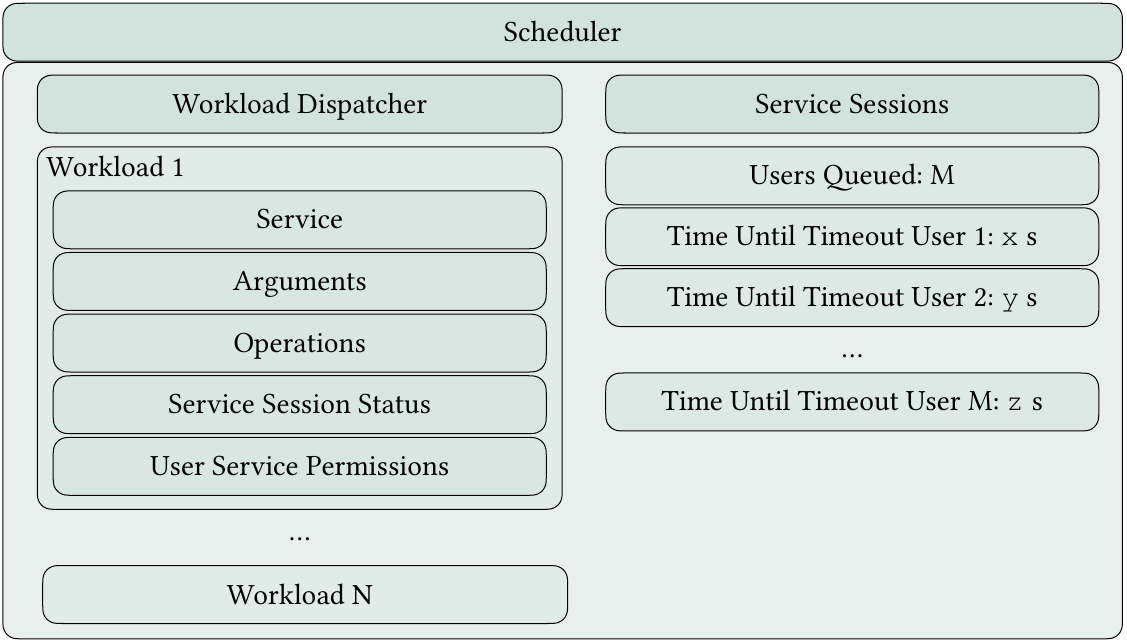}
    \caption{\texttt{Scheduler}. This component is responsible for assigning workloads to available workers (\texttt{Request Dispatcher}), based on metadata about the service session status and user service permissions (\texttt{Service Sessions}). This way, caching can be used more effectively, while the execution graph can receive all the necessary information for creating the workload.}
    \Description{}
    \label{fig:scheduler}
\end{figure}

Apart from the responsibilities of the \texttt{Scheduler} mentioned at Section~\ref{sec:scheduler_baseline_component}, this component includes additional functionality related to the integration of potential multiple services and other middleware components. This component is intended to schedule incoming requests (workloads) to available worker machines and keep track of the resources available in the system, with respect to active users. The workers can be in a pool of resources that can be accessed based on permissions. For example, we may want to restrict users' access to specialized hardware (GPUs) by default. The scheduler prioritizes the workload on a first-in-first-out basis (FIFO queue). In other words, requests that arrive at the scheduler are served on a first-come-first-serve basis. For every workload, the scheduler has information about the nature of the computation, such as what particular service it needs to call. Figure~\ref{fig:scheduler} shows a schematic overview of the scheduler. 

\subsection{\texttt{Service Identifier}}

This component is responsible for identifying services to be used in responding to user prompts. The \texttt{Service Identifier} is connected to a vector database to facilitate grounding predictions using information about available services (\texttt{Service\\Registry.get\_available\_services())}. Since the \texttt{Sched-\\uler} can access information about users' permissions, the task of excluding out-of-scope applications is handled separately. Consequently, the vector database is queried on a subset of data relevant to the query, ensuring compliance with user permissions. As a result, the \texttt{Service Identifier} generates a potential execution graph to solve the query — something that needs to be checked and parsed for correctness.

\begin{figure}[!b]
    \centering
    \includegraphics[width=0.4\textwidth]{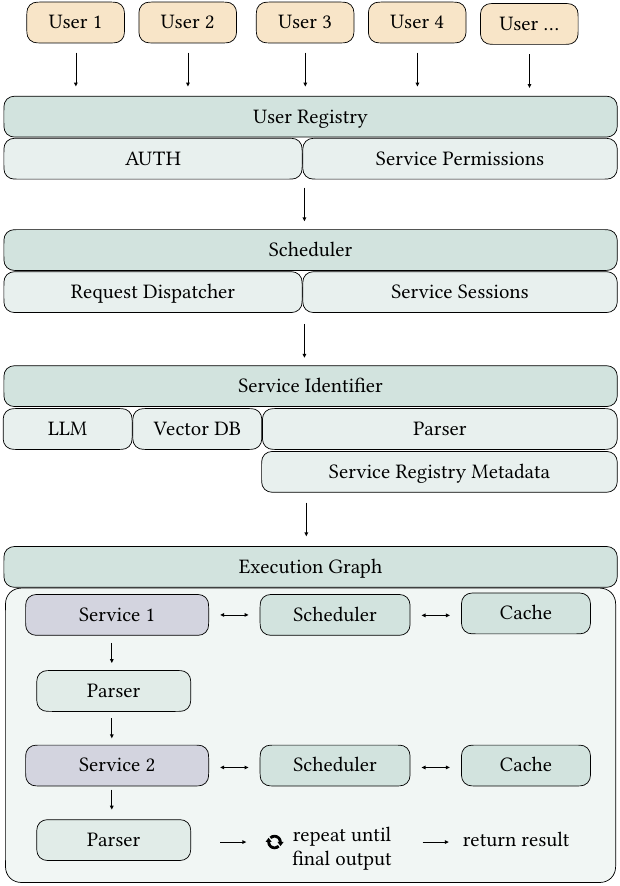}
    \caption{Example of end-to-end execution for LLMs as a gateway proposed middleware. Components in green are managed by the middleware, while components in other colors, such as a calculator service or a Python interpreter, are user-developed. Users are responsible for the logic of these applications and must ensure the input-output abstractions are correctly configured to interface with the middleware.}
    \Description{}
    \label{fig:end_to_end_execution_llm_app_using_middleware}
\end{figure}

\subsection{\texttt{Execution Graph}}
The \texttt{Execution Graph} component maintains a chain of services that loop until the initial prompt has been resolved (at the bottom of Figure~\ref{fig:end_to_end_execution_llm_app_using_middleware}). Inspired by TensorFlow's dataflow graph~\cite{199317}, this component ensures systematic processing. This component contains a chain of services that loop until the initial prompt has been solved. If additional information is needed from the user, the component will stop the execution process to obtain the required information and will resume execution once the user has provided the necessary input. Additionally, during the execution, caching per application is attempted for each service involved, with new computations executed only if the cache is unavailable. Additionally, this component includes a parser for each potentially intermediate output step to validate the calls made to the \texttt{Service Registry}.



\section{Prototype and Evaluation}

We implemented a conceptual prototype of our LLM middleware architecture in Python, focusing particularly on the service binding and protocol adaptation through the LLM. While the current implementation does not do its own scheduling and scaling, it models the user registry, service registry and service identifier components and allows new services to register with the middleware to extend the capabilities of the LLM. Additionally, we check the differences between a baseline LLM and our prototype when using a calculator to first showcase the benefits of augmenting LLM capabilities with conventional tools, and further we investigate how LLM responses scale with a varying number of arguments and a varying number of processes that share GPU resources. Lastly, we provide a forward-looking discussion on the scalability of LLM-based middleware, specifically regarding the \texttt{Execution Graph Generator}.

%

\subsection{Proof-of-concept: LLM with a Calculator Service}

We experiment with prompting an LLM to solve math questions typically handled by a calculator. To do this, we use two setups:
\begin{enumerate}
    \item \emph{LLM baseline}: we directly prompt the LLM to solve the math question
    \item \emph{LLM + Calculator Application}: we connect the LLM to a calculator application. The LLM here helps with identifying the operations and parameters needed to solve the question, and then it formulates the exact operations (add, subtract, etc.) and arguments (5, 5, 4, etc.) for the calculator.
\end{enumerate}

\begin{figure}[ht]
    \centering
    \includegraphics[width=0.48\textwidth]{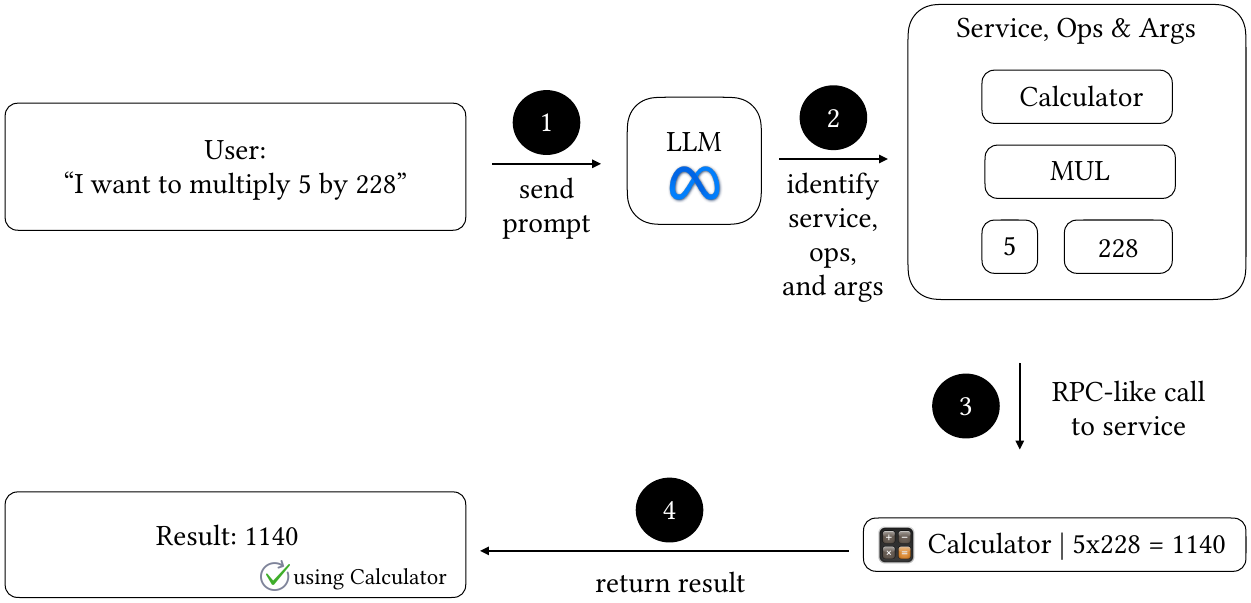}
    \caption{
\emph{LLM + Calculator Application}. Zero-shot prompting is used for querying the LLM — we prompt the model without potentially augmenting its capabilities using chain-of-thought reasoning or similar strategies.}
    \Description{}
\label{fig:workflow_LLM_app_calculator}
\end{figure}

Whereas for the former, we simply record the LLM responses and compare against ground truth samples, for the latter the workflow of Figure~\ref{fig:workflow_LLM_app_calculator} has been followed. We showcase the results in Table~\ref{tab:addition_use_case_1}. As it can be seen, the \emph{LLM + Calculator Application} significantly outperforms the \emph{LLM baseline}, especially as the number of arguments are increased, thus showcasing how accuracy can be significantly increased for certain types of tasks, just by connecting with an external application.

\begin{table}[b!]
  \centering
  \small
  \setlength\tabcolsep{1.5pt}  
  \begin{tabular}{|c|c|c|}
    \hline
    Args No.  & \textit{LLM baseline} accuracy & \textit{LLM + Calculator} accuracy \\
    \hline
    2 & 86/100 & 100/100 \\ 
    \hline
    3 & 37/100 & 100/100 \\ 
    \hline
    5 & 1/100 & 93/100 \\ 
    \hline
    10 & 0/100 & 97/100 \\ 
    \hline
    15 & 0/100 & 99/100 \\ 
    \hline
    20 & 0/100 & 99/100 \\
    \hline
     \end{tabular}
     \caption{Results for LLM with a Calculator Application. \textit{Args No.} here stands for the number of arguments present in the prompt. For example, a prompt has three arguments if it contains three numbers, e.g., \texttt{Add 5 to 3 to 2.} In this case, the arguments would be 5, 3, and 2. Accuracy in this context is computed by comparing against the ground truth expected results. For the LLM choice, we use LLama3~\cite{Meta_LLAMA}.}
       \label{tab:addition_use_case_1}
\end{table}

We observe that as we increase the number of arguments in the prompts, the LLM takes increasingly longer to calculate the response. For example, when the number of arguments is 2 (two numbers to be added), the LLM takes \texttt{1.031 s} to generate a response. As the number of arguments increases to 5, the response time grows to \texttt{1.685 s}. With 15 arguments and 20 arguments, the response times further increase to \texttt{2.235 s} and \texttt{3.108 s} respectively. Consequently, this indicates that as we increase the number of arguments in the query, the complexity increases accordingly, having a significant impact on the processing time of the LLM. Additionally, we test the response time of the LLM when we vary the number of processes that query it. Specifically, we notice that the response times are getting significantly lower when the GPU has to be shared. For instance, with 2 arguments, the response time increases from \texttt{1.031 s} to \texttt{2.058 s} when two processes share the GPU. Even worse, for the same query type with two arguments, when three processes share the GPU, the response time increases from \texttt{1.031 s} to \texttt{8.548 s}. This further highlights the importance of resource management and ability to scale accordingly in scenarios where multiple processes might be accessing an LLM simultaneously.


\subsection{Scaling of LLM Execution Graph Generator}
As the \texttt{Execution Graph} is the key component for processing user prompts and returning results, its scalability is of critical importance. Specifically, in future work, we aim to explore its throughput, in scenarios where we vary the number of services available in the framework. For example, it would be interesting to determine the performance of this component when the \texttt{Service Registry} contains only one service, as well as when it includes a larger number of services. Additionally, to ensure the feasibility of using this component, it would be important to investigate the potential impact of varying the amount of information appended to the query, such as providing either more thorough or shorter descriptions for the services. This is especially relevant if the \texttt{Service Identifier} has to decide what services to use based on this information. Moreover, understanding the performance of using a vector database in this context would also be critical. For instance, as some services might rely on grounding their responses in documents (for LLM-based services), filtering the vector database before querying it might yield performance advantages. Specifically, for answering queries, different parts of the database can be ignored by design, given that different users have access to only subsets of the database space due to varying service permissions.

\section{Related Work}
\label{sec:related_work}

While none of the existing approaches come close to our vision of a comprehensive middleware for LLMs, several fragmented approaches exist enabling the integration of LLMs into larger systems. 

\textbf{LangChain}~\cite{topsakal2023creating} is a framework enabling the development of LLM-based solutions. As the name connotes, the built LLM solutions can consist of several chains of microservices.
The applications can access additional data sources within the deployed environment using RAG pipelines. LangChain also utilizes agents for internal reasoning based on ReAct (Section~\ref{subsec:Explainability}). Thus, it can access external knowledge over the Internet. Users can interact with LangChain in a chat via prompts~\cite{LangChain_Concept}.
LangChain also implements other functionalities, like LangSmith to enable AI Observability: LangSmith allows users to trace the LLM pipeline starting. The trace can contain information on the user prompt, the retrieved model output, and further metadata. All collected traces are also available in a monitoring tool allowing for visual analytics \cite{LangSmith_Observability}. In comparison, our vision enables self-hosting LLMs on-premise, ensuring data privacy while extending beyond Lang Chain's functionality by integrating and orchestrating multiple LLMs. Unlike LangChain, which focuses on chaining services mainly around a single LLM, we envision a system that facilitates interactions not only between multiple LLMs and users but also among the LLMs themselves. This approach addresses a more complex problem, enabling the efficient use of multiple LLMs at scale and on premise, integrating external applications, and potentially incorporating LLMs as part of the middleware.


\textbf{PrivateGPT} has been created to allow the usage of generative AI for privacy-related documents \cite{Martinez_Toro_PrivateGPT_2023}. It implements a RAG pipeline and exposes APIs for document ingestion. 
PrivateGPT offers three different usage modes that must be selected before interacting with the LLM with prompts. The first mode is a simple, non-contextual LLM chat. Only previous chat messages but not ingested documents are considered in this mode. In another mode, users can ask the LLM questions about already ingested documents. Here, previous chat messages are also taken into account. The last mode offers a search within the ingested documents. The search returns the four most relevant text chunks including the source documents. 
PrivateGPT is restricted in its usage by only interacting with the user and their ingested documents but not with other systems. The different modes limit its functionality.

Amazon offers a similar service with \textbf{Amazon Bedrock}. It enables the integration of foundation models (FM) into the user's application using AWS tools. 
Bedrock allows the customization of AI models and implements RAG. It can also interact with the enterprise system and data sources using \textbf{agents}. Agents fill the bridging gap between the FM, the user, and the user's end-system. They are responsible for understanding the user request to break it down into smaller steps. Based on that, they perform API calls to the end system. Agents can re-prompt the user or query data from a knowledge base if any information is missing. They retrieve the API call results and invoke the FM to understand the output. Afterwards, agents evaluate whether further steps or information is needed, or if the result is sufficient. Depending on the decision, they reiterate the process of doing more API calls and prompting the model, re-prompts, or returning a result to the user \cite{Amazon_Bedrock_Agents}.
Similarly to LangChain, Amazon Bedrock focuses on embedding LLMs into the ecosystem. However, compared to our vision, Bedrock is cloud-based, while our solution can be run on-premise ensuring data privacy. Further, Bedrock does not facilitate chaining up LLMs to enable LLM-to-LLM communication. Bedrock's tracing feature captures all metadata within logs at each possible step. Nevertheless, currently, there is no functionality to detect whether drifts within the data or the model happened which is part of our vision.

\noindent \textbf{Seldon} and \textbf{NVidia} offer a different software solution. Rather than using LLMs for service and task orchestration, they provide functionalities to integrate AI models into the environment as microservices \cite{NVIDIA_Triton, Seldon_Core}. 
They are deployed into given environments. NVidia provides an inference server for the models, whilst Seldon employs a Kubernetes cluster. The environment manages, orchestrates, and observes deployed microservices. Further, both platforms focus on integrating different ML models and boosting the system's efficiency. They are not directly taking ML microservices into account for inter-microservice communication, thus, not considering the usage of LLMs within microservice chains. 

\textbf{Amazon Alexa} can be seen as an early example of an extensible agent system that shows properties similar to our proposed LLM as a Gateway design. It integrates classically programmed services and a natural language user interface. In this regard, it can serve as a gateway to Internet services for home users and aggregate information and services in the manner of a personal assistant. Service developers provide a specifically shaped interface (skill) for the natural language component to latch onto, and the natural language component identifies the skill to use and parameters to pass into it. Each skill bundles a number of intents, each of which is effectively a function call: it has a name and an arbitrary number of typed slots, performs some actions and returns a result that the natural language component renders into natural language.  For intents that require many slots, confirmation or other multi-turn scenarios, there is a dialogue manager~\cite{Amazon_dialogue}
that can be classically programmed or the  Conversations framework~\cite{Alexa_Conversations},
in which a language model is trained on utterances and example conversations to keep track of the conversation and extract intent and slot values over
multiple prompts.





\section{Conclusions and Future Research}
The increasing interest in integrating LLMs within enterprise environments proves their tremendous potential in enhancing user interfaces, search capabilities, and analytics tasks. While commercial cloud-based solutions like OpenAI's ChatGPT or Anthropic's Claude have led the way, there is a notable gap in the availability of on-premise LLM solutions, particularly those that provide robust middleware support. 

In this work, we have outlined our vision for a comprehensive middleware, discussed where hosting and integrating LLMs introduces challenges beyond traditional software components, but also shown ways in which LLMs can become an active part of a middleware to address these challenges. 

Many aspects of our vision, e.g., effective scheduling of multiple versions of big LLMs or the observability challenge, are touching on problems that have significant research gaps or areas  in which theoretical results have not been successfully implemented in practice. We therefore consider the problem of designing and building middleware for LLMs to become a longer research effort for our community.


\bibliographystyle{plain}


\end{document}